\documentstyle[revtex,eqsecnum]{aps}

\tightenlines
\begin{document}
\draft

\preprint{DOE/ER/40322-163, U. of MD PP \#92-225}

\begin{title}
Response of nucleons to external probes in hedgehog models: \\
II. General formalism
\end{title}

\author{Wojciech Broniowski \cite{ifj} and Thomas D.  Cohen}
\begin{instit}
Department of Physics and Astronomy, University of Maryland \\
College Park, Maryland 20742-4111
\end{instit}

\begin{abstract}
Linear response theory for SU(2) hedgehog
soliton models is developed in analogy
to a standard method in many-body physics.
In this framework, we discuss response of baryons to external
probes, and develop expressions for polarizabilities.  We discuss isospin
effects (neutron-proton splitting) in polarizabilities. Methods
for cases with zero modes are presented, including
numerical techniques. Our approach is
based on the $1 / N_c$-expansion scheme. We work
in a model with quark and meson degrees of freedom, but
the basic method is valid in any hedgehog model, such as the Skyrmion or
the Nambu--Jona-Lasinio model in the solitonic treatment. The equations of
motion for coupled RPA quark-meson fluctuations are classified
according the hedgehog symmetries, and are written down explicitly in the
grand-spin basis.
\end{abstract}


\narrowtext
\section{Introduction}
\label{se:introduction}

In recent years various hedgehog models (Chiral Quark Meson (CQM) models
\cite{BirBan8485,KRS84,KR84,EisKal,BBC:rev,MCB:rev},
Skyrme models \cite{Skyrme6162,ANW83,Adkins84,skyrmion:rev},
hybrid bag models \cite{hybrid:rev},
chiral models with confinement
\cite{BBC:rev,MKB:conf,BrLj,Duck}
or the
Nambu--Jona-Lasinio (NJL) model \cite{NJL}  in the solitonic treatment
\cite{Dyakonov86,DPPob88,MAGGM,Reihardt88,Meissner89,DPPr89,Alkofer})
were extensively applied to describe the physics of low-energy baryons.
Semiclassical methods for treatment of these
models, such as various projection
methods \cite{CB86,GolliRosina,Birse85,RenBan}, or
RPA method \cite{BC86} were developed. Masses, various
charges, $\pi-N$ phase shifts \cite{piN}, were
calculated, with quite reasonable agreement
with experiment, depending on the specific model, number of
included fields, etc. In this article we develop the linear response
formalism for hedgehog models. We work in the
framework of a CQM model, since it
has both quark and meson degrees of freedom,
and in this respect has the essential
features of both the purely mesonic Skyrme model, and the
NJL model, which involves quark degrees of
freedom only. Our methods and final
expressions can be modified straightforwardly
to be applicable in these models.

Hedgehog models can be used to describe response of nucleons to
external probes, and to calculate corresponding
polarizabilities. A natural approach is the
linear response method
of many-body physics \cite{Ring}. The underlying picture is as follows:
A current interacts with the nucleon, creates an
intermediate state which is an RPA phonon
excitation on top of the soliton.  This state
interacts with another current and de-excites back into a nucleon state.
The RPA phonon states are constructed from
one-particle--one-hole excitations
of the quarks, as well as from quantum meson
excitations. Quark and meson
fluctuations are coupled, and the resulting
equations of motion for the fluctuations are solved.
An example of physically important two-current
observables which can be
calculated in this way are the electromagnetic polarizabilities
of the nucleon \cite{BBC91nt}.
This topic is extensively studied in the preceding paper \cite{part2},
henceforth referred as (I). The present article is devoted to development
of the necessary formalism, and contains many technical but
necessary details of linear response in hedgehog models.
while (I) concentrates on physical aspects.

This article is organized as follows:
In Sec. \ref{se:hedgehogs} we very briefly review
a CQM model \cite{BB8586}, its soliton solutions (Sec. \ref{se:solitons}),
as well as hedgehog symmetries (Sec. \ref{se:symmetries}).
One of the discrete symmetries, the grand-reversal symmetry
\cite{CohenBan86,CB86}, will be
particularly useful in classifying various
perturbations. Section \ref{se:linear}
is the core of the paper, and describes the
equations-of-motion approach \cite{Rowe} to
linear response in hedgehog models. We start
from deriving small-fluctuation equations
of motion (Sec. \ref{se:eom}) for coupled
quark-meson systems driven by an
external perturbation. These equations are classified according to
hedgehog symmetries. In the static limit the
grand reversal symmetry, ${\cal R}$, decouples
the equations into odd-${\cal R}$ equations,
involving quarks only, and even-${\cal R}$ equations,
involving both quarks and mesons. We discuss
in detail the problem of zero modes
(Sec. \ref{se:zeromodes}). These zero modes arise from
braking of the symmetries of the lagrangian
by the soliton solution. In our
applications, we will have to deal with
rotational zero modes (cranking)
and translational zero modes (isoscalar
electric perturbation in (I)). We describe
a numerical method to deal with the excitation
of zero modes whose amplitude diverges as the
frequency of the perturbation goes to zero
(Sec. \ref{se:numerical}). We  discuss
stability of solitons (Sec. \ref{se:stabil}).
In Sec. \ref{se:crank} we present cranking in the
linear response  formalism. Quantization via cranking
is reviewed in  Sec. \ref{se:quant}.
In  Sec. \ref{se:external} we describe
the calculation of polarizabilities
in states of good spin and isospin, and
obtain our basic formulas. Section
\ref{se:examples} illustrates the method by
presenting the standard calculation
of the $N$-$\Delta$ mass splitting, as well as
the evaluation of the neutron-proton hadronic
mass difference. The issues
of $N_c$-counting are discussed in
Sec. \ref{se:nc} . We show how to apply
the linear response in a way consistent
with $1 /N_c$-expansion scheme.
Finally, Sec. \ref{se:other} contains remarks
relevant to other models (Skyrme model, NJL).

Appendices contain some details of the grand-spin
algebra, derivation of the explicit
forms of the equations of motion for
fluctuations in of the grand-spin basis
(App. \ref{ap:eqmotion}), and a glossary
of useful formulas with collective
matrix elements (App. \ref{ap:collective}).
We also give a simple proof of equality of the soliton mass
and the inertial mass parameter (App. \ref{ap:masses}),
and discuss the issue of Pauli blocking of the Dirac sea
in chiral quark models (Sec. \ref{ap:Pauli}).

\section{Hedgehog models}
\label{se:hedgehogs}

In this paper most of the derivations
will be done in the framework of the
chiral quark-meson model (CQM) of Ref. \cite{BirBan8485}.
For the details, description of solutions, and
the resulting phenomenology obtained
with the cranking projection method,
the reader is referred to Ref. \cite{CB86}.
The reason of choosing
this particular model over other models,
e.g. the Skyrmion or the NJL model, is that
it contains both quark and meson
degrees of freedom, and formally has
all essential features of a generic
hedgehog model with two flavors.
At the same time, it is free of the
non-linear complications of the Skyrme model,
or the Dirac-sea complications of the NJL model.

\subsection{Soliton solutions}
\label{se:solitons}

The lagrangian of the model is the
Gell-Mann--L\'evy lagrangian
\cite{GML}, with $\psi$ denoting the
quark operator, and $\sigma$ and $\mbox{\boldmath $\pi$}$
denoting the meson fields:
\begin{eqnarray}
 {\cal L} &=& \bar{\psi} \left [
\dot{\imath} \overlay{\slash}{\partial}
   + g \left ( \sigma+\dot{\imath}
{\gamma_{5}} \mbox{\boldmath $\tau$} \cdot
\mbox{\boldmath $\pi$} \right ) \right ]
\psi \nonumber \\
    &+&  \mbox{$1\over 2$}
(\partial{}^{\mu} \sigma)^{2} +
\mbox{$1\over 2$} (\partial{}^{\mu} \mbox{\boldmath $\pi$})^{2}
   - U \left ( \sigma,
\mbox{\boldmath $\pi$} \right ).   \label{eq:GML}
\end{eqnarray}
The Mexican Hat potential,
\begin{eqnarray}
&&U \left ( \sigma, \mbox{\boldmath $\pi$} \right ) =
    \frac{\lambda^2}{4} {\left (
\sigma^{2} + \mbox{\boldmath $\pi$}^{2} -
                \nu^2 \right )}^{2} \nonumber +
{m_\pi}^{2} {F_\pi}\sigma , \nonumber \\
&&\lambda^2 = \frac {{m_\sigma}^{2} -
{m_\pi}^{2}} {2 {F_\pi}^{2}} , \;\;\;
\nu^2 =     \frac{{m_\sigma}^{2} - 3 {m_\pi}^{2}}
                  {{m_\sigma}^{2} -   {m_\pi}^{2}} ,
\label{eq:Potential}
\end{eqnarray}
leads to the spontaneous breaking of the chiral symmetry in the
usual way \cite{GML,BirBan8485}. Our
convention for the pion decay constant
is ${F_\pi} = 93 MeV$.
At the (time-dependent) mean-field level,
only valence quarks, denoted by $q$,
are retained in the expansion of the quark
fields, and the meson fields are treated as
classical, c-number fields \cite{CB86}
(see also App. \ref{ap:Pauli}). The time-dependent equations
of motion have the form
\begin{eqnarray}
&& (h[\phi] - \dot{\imath} \partial_{t}) q = 0 ,  \\
&& - \Box \phi = \frac {\delta U[\phi]}{\delta \phi}
   - g N_c \bar{q}{M} q, \label{eq:timedependent}
\end{eqnarray}
where $\phi = (\sigma, \mbox{\boldmath $\pi$})$
denotes the meson fields,
${M} = (\beta, \dot{\imath} \gamma_{5}
\mbox{\boldmath $\tau$})$ describes the quark-meson
coupling, and the Dirac hamiltonian is
$h[\phi] = - \dot{\imath} \mbox{\boldmath $\alpha$}
\cdot \mbox{\boldmath $\nabla$} - g {M} \phi$.
Equations (\ref{eq:timedependent}) have
a stationary solution of the form
\FL
\begin{eqnarray}
&& \sigma(\mbox{\boldmath $r$},t) = {\sigma_h}(r), \;\;\;
   \mbox{\boldmath $\pi$}(\mbox{\boldmath $r$},t) =
\widehat{\mbox{\boldmath $r$}} {\pi_h}(r), \;\;\;
   q(t) = q_h (\mbox{\boldmath $r$})
e^{-\dot{\imath} \varepsilon t}, \nonumber \\
&& q_h = \left ( \begin{array}{c}
               G_h(r) \\
               \dot{\imath} \mbox{\boldmath $\sigma$}
\cdot \widehat{\mbox{\boldmath $r$}} F_h(r)
               \end{array}
        \right )
         (|u \downarrow \rangle - |d \uparrow \rangle)/\sqrt{2},
\label{eq:stationarysol}
\end{eqnarray}
where $\varepsilon$ is the quark eigenvalue.
For discussion of this solution,
plots of the radial functions $\sigma_h$,
$\pi_h$, $G_h$, and $F_h$,
and other details,  the reader is referred to
Refs. \cite{BirBan8485,CB86}.

\subsection{Hedgehog symmetries}
\label{se:symmetries}

The solution (\ref{eq:stationarysol}) has the hedgehog form,
which breaks the spin, $\mbox{\boldmath $J$}$,
and isospin, $\mbox{\boldmath $I$}$, symmetries
of the lagrangian (\ref{eq:GML}),
leaving as a good symmetry the grand
spin, $K = I + J$. There are also two discrete symmetries which are
very useful in classifying solutions and perturbations.
One is parity, ${\cal P}$, the other is the ``grand-reversal'' symmetry,
${\cal R}$, discussed in Refs. \cite{CohenBan86,BC86}.
Formally, ${\cal R}$ is defined as the time-reversal, followed by an
isorotation by angle $\pi$ about the 2-axis in isospin. Explicitly,
it transforms the quark spinors and
mean meson fields as follows \cite{CB86}:
\begin{eqnarray}
&& q(\mbox{\boldmath $r$},t) \rightarrow
\sigma_z \tau_2 q^{*}(\mbox{\boldmath $r$},-t),
\nonumber \\
&& \sigma(\mbox{\boldmath $r$},t) \rightarrow
\sigma^{*}(\mbox{\boldmath $r$},-t), \;\;\;
   \pi(\mbox{\boldmath $r$},t) \rightarrow
\pi^{*}(\mbox{\boldmath $r$},-t).
    \label{eq:grandreversal}
\end{eqnarray}
We denote the action of ${\cal R}$ on an
object by the superscript~${}^{\cal R}$.
The soliton solution has $K^{\cal PR} = 0^{++}$.

\section{Linear response in hedgehog models}
\label{se:linear}

In this section the basic formalism of
linear response in hedgehog models
is developed. We use the equation-of-motion approach \cite{Rowe},
which is based on solving equations of motion
for small oscillation on top of the
ground state solutions. This method is equivalent to the
particle-hole formalism \cite{Ring}, in which one introduces
a quantum RPA state, quasi-boson RPA phonon operators, etc. Methods
such as cranking, projection, or
quantization of zero modes, can be described
in this framework, and have
definite quantum-mechanical interpretation.
For simplicity of notation, we present our
formalism in the equations-of-motion method.

\subsection{Equations of motion for small fluctuations}
\label{se:eom}

Let us consider a small oscillation problem in our system.
We introduce shifts in the valence quark
spinor and in the meson fields,
\begin{eqnarray}
&& \delta q(\mbox{\boldmath $r$},t) =
\left ( X(\mbox{\boldmath $r$}) e^{- \dot{\imath}
\omega t} + Y^{\cal R}(\mbox{\boldmath $r$})
e^{\dot{\imath} \omega t} \right )
e^{- \dot{\imath} \varepsilon t}, \nonumber \\
&& \delta \phi_{a}(\mbox{\boldmath $r$},t) =
Z_{a}(\mbox{\boldmath $r$}) e^{- \dot{\imath} \omega t} +
Z_{a}^{\cal R}(\mbox{\boldmath $r$})
e^{ \dot{\imath} \omega t},
\label{eq:RPAansatz}
\end{eqnarray}
where $X$ and $Y$ describe the shift in the valence quark
spinor, and $\delta \phi_{0}$ and
$\delta \mbox{\boldmath $\phi$}$ are the shifts in the
$\sigma$ and $\mbox{\boldmath $\pi$}$
fields, respectively. Note, that in
Eqs. (\ref{eq:RPAansatz}) the ${\cal R}$ transformation has taken
the place of the usual \cite{Ring} complex conjugation. This is
because in hedgehog systems the grand-reversal replaces the usual
time-reversal symmetry. According to
definition (\ref{eq:grandreversal}), the meson shifts $\delta \phi$ are
even under grand-reversal, but the quark shifts have in general both
even and odd components. We
linearize equations (\ref{eq:timedependent})
about the solitonic solution
(\ref{eq:stationarysol}), and obtain the quark-meson RPA equations. When
external perturbations are present, these equations are in general
driven by a quark source, $j_q$, and a meson source, $j_{\phi}$,
\begin{eqnarray}
j_q &=& j_X e^{- \dot{\imath} \omega t} +
j_Y^{\cal R} e^{\dot{\imath} \omega t},
\nonumber \\
j_{\phi} &=& j_{Z} e^{- \dot{\imath} \omega t} +
j_{Z}^{\cal R} e^{\dot{\imath} \omega t},
\label{eq:sources}
\end{eqnarray}
Again, the meson source is even under
${\cal R}$, whereas the quark source has
in general even  and odd-${\cal R}$ components.
Using the fact that $h[\phi_h]$,
$M$ and $q_h$ are even under ${\cal R}$ (in
fact they are $K^{\cal PR} = 0^{++}$
objects), we obtain a general form of the linear response equations
for our hedgehog system:
\begin{eqnarray}
&& \left ( h[\phi_h] - \varepsilon  \right ) X
  - g \sum_{a} M_{a} q_{h} Z_{a} -
\omega X  =  j_X, \nonumber \\
&& \left ( h[\phi_h] - \varepsilon  \right ) Y
- g \sum_{a} M_{a} q_{h} {Z}_{a} +
\omega Y  =  j_Y, \nonumber \\
&& -\nabla^{2} Z_{a} + \sum_{b}
\left . \frac{\delta^{2} U}{\delta
\phi_{a} \delta \phi_{b}}
 \right |_{\phi = \phi_{h}} Z_{b}  \nonumber \\
&& \hspace{0.2in} - g N_c \left (
q_{h}^{\dagger} M_{a} X +
Y^{\dagger} M_{a} q_{h} \right ) -
\omega^{2} Z_{a}  =  j_{Z}.
\label{eq:lineareq}
\end{eqnarray}
Introducing auxiliary meson momentum
variables $P_a = - \dot{\imath} \omega Z_{a}$, we observe that
Eqs. (\ref{eq:lineareq}) can be
written in the symplectic form \cite{Ring}
\begin{equation}
{\cal H} \xi - \omega \Lambda \xi = j,
\label{eq:linear}
\end{equation}
where ${\cal H}$ is the RPA hamiltonian, and $\Lambda$ is the
symplectic RPA metric, satisfying $\Lambda^2 = 1$.
In the grand-spin
basis (App. \ref{ap:eqmotion}), ${\cal H}$ is real.
Our problem (\ref{eq:lineareq}) can then be written as
\begin{equation}
{\cal H} = \left ( \begin{array}{cccc}
 N_c (h - \varepsilon)     &     0
 & - g N_c M q_h        &   0 \\
                  0        &  N_c (h   - \varepsilon)
 & - g N_c M q_h        &   0 \\
 - g N_c q_h^{\dagger} M   &  - g N_c q_h^{\dagger} M
 & - \nabla^2 + U''     &   0 \\
 0                         &       0
 &               0      &   1
                   \end{array} \right ) ,
\label{eq:H}
\end{equation}
\begin{equation}
\Lambda = \left ( \begin{array}{cccc}
  1     &     0   &  0  &   0 \\
  0     &     -1  &  0  &   0 \\
  0     &     0   &  0  & \dot{\imath} \\
  0     &     0   & -\dot{\imath} &   0
                   \end{array} \right ) ,
\;\;\; \xi = \left ( \begin{array}{c}
  X  \\
  Y  \\
  Z  \\
  P
                   \end{array} \right ) ,
\;\;\; j = \left ( \begin{array}{c}
  N_c j_X  \\
  N_c j_Y  \\
  j_Z  \\
  j_P
                   \end{array} \right ) .
\label{eq:Lxi}
\end{equation}
Note the appearance of an odd-${\cal R}$ momentum component
in the source, $j_P$, which arises in some cases (cranking).

We are interested in the limit of vanishing frequency of the external
perturbation, $\omega \rightarrow 0$. If zero modes are excited by an
even-${\cal R}$ perturbation (Sec. \ref{se:zeromodes}),
then the full equations
Eq. (\ref{eq:lineareq}) have to be solved. Otherwise, one can set
$\omega = 0$ and deal with simplified cases.  At this
point the grand-reversal
classification becomes very useful. Acting with ${\cal R}$
on Eq. (\ref{eq:linear}) effectively replaces $X \leftrightarrow Y$,
$j_X \leftrightarrow j_Y$, $Z \leftrightarrow Z$,
$j_Z \leftrightarrow j_Z$,
$P - \leftrightarrow P$, and $j_P - \leftrightarrow j_P$. Let us
introduce odd and even grand-reversal combinations:
$\delta q^{\pm} = X \pm Y$, $j^{\pm}_q = j_X \pm j_Y$, and rewrite
Eq. (\ref{eq:linear}) by adding and subtracting the first two
equations. We get for the case of an odd-${\cal R}$ perturbation
\begin{eqnarray}
(h - \varepsilon) \delta q^{-} &=& j^{-}_q , \nonumber \\
P &=& j_P ,
\label{eq:oddR}
\end{eqnarray}
and for the case of an even-${\cal R}$ perturbation
\begin{eqnarray}
(h - \varepsilon) \delta q^{+} - 2 g M q_h Z &=& j^{+}_q ,
\nonumber \\
(- \nabla^2 + U'') Z - N_c g q_h^{\dagger} M \delta q^{+} &=& j^Z
\label{eq:evenR}
\end{eqnarray}
The odd-${\cal R}$ equations (\ref{eq:oddR})
involve a quark field equation,
and a trivial equation for $P$. The even-${\cal R}$
equations (\ref{eq:evenR})
involve coupled quark and meson fluctuations.
Equations (\ref{eq:linear}), or
(\ref{eq:oddR},\ref{eq:evenR}) are further decomposed
by grand-spin, $K$, and parity, ${\cal P}$
(App. \ref{ap:eqmotion}).

In models with vector mesons, such as \cite{BB8586}, the odd-${\cal R}$
equations may also involve mesonic shifts. For example,  the space
components of the $\omega$ meson and the time componenent of the $\rho$
meson enter into the cranking equations of motion \cite{BCVector}.

\subsection{Zero modes}
\label{se:zeromodes}

First consider the undriven problem (\ref{eq:linear}),
with $j = 0$, which determines the
RPA spectrum and eigenmodes. A complication arises
whenever a continuous symmetry of
the lagrangian is broken by the solitonic solution,
e.g. translation, or
rotational symmetry. For each broken symmetry the small
fluctuation equations have a pair of
zero-modes \cite{Ring}: $\xi_0$, the symmetry mode, obtained
by acting with a symmetry generator on the
solitonic solution, and
a conjugate zero mode, $\xi_1$. They satisfy the equations
\begin{eqnarray}
{\cal H} \xi_0 & = & 0, \\
{\cal H} \xi_1 & = & - \dot{\imath} \Lambda \xi_0.
\label{eq:zero}
\end{eqnarray}
The remaining ``physical'' modes, $\xi_i$,
satisfy the equations
\begin{equation}
{\cal H} \xi_i = \omega_i \Lambda \xi_i .
\label{eq:ap:freelinear}
\end{equation}

One can easily show that the symplectic norms satisfy conditions
\begin{eqnarray}
&&\xi_0^{\dagger} \Lambda \xi_0 = \xi_1^{\dagger}
\Lambda \xi_1 = 0, \;\;\;
\xi_0^{\dagger} \Lambda \xi_1 = - \mbox{${\dot{\imath}}\over 4$}
{\cal M}, \nonumber \\
&&\xi_i^{\dagger} \Lambda \xi_j = \mbox{$1\over 4$}
\delta_{ij} {\cal N}_i , \nonumber \\
&&\xi_0^{\dagger} \Lambda \xi_i =
\xi_1^{\dagger} \Lambda \xi_i = 0, \;\;\; i = 2,3,...
\label{eq:norms}
\end{eqnarray}
where ${\cal M}$ is the appropriate inertia
parameter (mass, moment of inertia)
parameter, and ${\cal N}_i$ are the symplectic
norms of the physical modes.
The factors of $\mbox{$1\over 4$}$ are conventional,
and factors of $\dot{\imath}$ are inserted for convenience.
Expanding the solution of Eq. (\ref{eq:linear})
in RPA eigenmodes,
\begin{equation}
\xi = \sum_{\mu=0,1,2,...} c_{\mu} \xi_{\mu},
\label{eq:expand}
\end{equation}
introducing ``charges'':
$Q_{0} =  4 \dot{\imath} \xi_{0}^{\dagger} j$,
$Q_{\mu} = 4 \xi_{\mu}^{\dagger} j$, $\mu = 1,2,...$, and using
Eqs. (\ref{eq:norms}), we find that
\begin{eqnarray}
&&\omega c_1 {\cal M} + Q_0 = 0, \;\;\;
\dot{\imath} \omega c_0 {\cal M} + Q_1 - c_1 {\cal M} = 0,
\nonumber \\
&& c_i ({\omega}_i - \omega) {\cal N}_i = Q_i .
\label{eq:coeffrelations}
\end{eqnarray}
We consider two cases which arise in practical
applications: 1) $Q_1 =0$, and 2) $Q_1 \neq 0$, $Q_0 = 0$.

\subsubsection{Case $Q_1 = 0$}
\label{se:lineara}

Using Eq. (\ref{eq:coeffrelations}) we find
\begin{equation}
c_0 =  \frac{\dot{\imath} Q_0}{{\cal M} \omega^2}, \;\;\;
c_1 = - \frac{Q_0}{{\cal M} \omega}, \;\;\;
c_i =   \frac{Q_i}{{\cal N}_i (\omega_i - \omega)} .
\label{eq:coeffc}
\end{equation}
The second-order energy shift, $\kappa$,
corresponding to a given perturbation
(a ``polarizability'' is equal to $2 \kappa$)
is given by the usual perturbation theory result
\begin{eqnarray}
&&\kappa =  2 \xi^{\dagger} j = \sum_{\mu} c_{\mu}^{*} Q_{\mu} =
\kappa^{zero} + \kappa^{phys.}, \nonumber \\
&&\kappa^{zero} = - \mbox{$1\over 2$}
\frac{Q_0^2}{{\cal M} \omega^2}, \;\;\;
\kappa^{phys} = \mbox{$1\over 2$} \sum_{i}
\frac{Q_i^2}{{\cal N}_i (\omega_i - \omega)}.
\label{eq:polarizability}
\end{eqnarray}
In the limit $\omega \rightarrow 0$, the coefficients $c_{0}$, $c_{1}$ and
the zero-mode part of $\kappa$ diverge, as long as $Q_0 \neq 0$. This
has a physical interpretation: for instance in the case of translation
the center of mass of the system moves, and the amplitude of this motion,
$c_{0}$, as well as ``velocity'', $c_{1}$, diverge. In (I) we show
how this feature of the linear response formalism
leads to the Thompson limit of the Compton scattering amplitude.

\subsubsection{Case $Q_1 \neq 0$, $Q_0 = 0$}
\label{se:linearb}

In this case we can take the limit $\omega \rightarrow 0$
on the outset, and from
Eq. (\ref{eq:coeffrelations}) we get
\begin{equation}
c_1 =  \frac{Q_1}{{\cal M}} , \;\;\;
c_i =   \frac{Q_i}{{\cal N}_i \omega_i} .
\label{eq:coeff1}
\end{equation}
The amplitude of the symmetry mode, $c_0$, remains undermined.
The second-order
energy shift is:
\begin{equation}
\kappa = \mbox{$1\over 2$} \frac{Q_1^2}{{\cal M}} +
\mbox{$1\over 2$} \sum_{i} \frac{Q_i^2}{{\cal N}_i \omega_i}.
\label{eq:polarizability1}
\end{equation}

\subsection{Numerical methods in presence of diverging zero modes}
\label{se:numerical}

Numerically, the excitation of amplitude-growing zero
modes (Sec. \ref{se:lineara}) creates special difficulties in
extracting the ``physical'' parts of observables, e.g.
electromagnetic polarizabilities. The problem can be remedied
as follows: We solve Eqs. (\ref{eq:linear}) for a small value of
$\omega$. Next, we project out the zero-mode part from $\xi$, obtaining
$\xi^{phys.} = \xi - c_0 \xi_0$, and calculate physical parts of
observables. The procedure is repeated with decreasing $\omega$, until
the results no longer change. In practice, a very high accuracy of
the soliton solution, as well as the fluctuation solutions, is required
for this procedure to be feasable. A better method is to
project the part of the source, $j$, which couples to the zero mode,
and solve equations
\begin{equation}
{\cal H} \xi^{phys.} - \omega \Lambda \xi^{phys.} = j^{phys.},
\label{eq:projectj}
\end{equation}
where $j^{phys.} = j - ({Q_0}/{{\cal M}}) \Lambda \xi_1$,
and $\xi_1$ is obtained
by solving Eq. (\ref{eq:zero}) first.
Equations (\ref{eq:projectj}) do
not excite the zero mode, and directly lead to the
physical part of the solution. The advantage of
the method with the projected source over
the direct method described
previously follows from the fact that in numerical
solutions to Eqs. (\ref{eq:projectj}) the admixtures
of the zero mode arise only from numerical noise. Their
amplitude is small, such that we can easily
control numerical precision
in the physical mode. Because of these admixtures,
a small nonzero value of $\omega$ should
be kept as a regulator in
Eqs. (\ref{eq:projectj}), and the zero-mode contamination
has to be projected out after the numerical solution is found.

\subsection{Stability of solitons}
\label{se:stabil}

Since in our problem ${\cal H}$ and $\Lambda$ are hermitian,
one finds that ${\cal H}^2 \xi_i = \omega_i^2 \xi_i$
is a hermitian
eigenvalue problem. Therefore in our case
$\omega_i^2$ are real, and $\omega_i$
can either be purely real, or purely imaginary.
The modes appear in
conjugated pairs $(\xi_i,\xi_j)$, with $\omega_i = - \omega_j$.
If the spectrum
contains an imaginary eigenvalue, we have to instability
(in the Lyapunov sense
\cite{Arnold}) of the ground-state (soliton) solution \cite{FLS,JJR}, and
of course linear response on top of an unstable system makes no sense.
In Ref. \cite{BC86} we have shown that the soliton of
Ref. \cite{BirBan8485} is stable with
respect to breathing modes, i.e. the $K^P = 0^+$
excitations. With the explicit forms of the equations
in App. \ref{ap:eqmotion}, stability could be checked numerically
for any $K^P$ vibrational mode. It is generally believed that the hedgehog
solitons are indeed stable, although it has not been proved analytically or
numerically.

\subsection{Cranking as linear response}
\label{se:crank}

Cranking \cite{CB86} may be viewed as linear response.
In a frame iso-rotating with a small angular velocity
$\mbox{\boldmath $\lambda$}$, we discover equations
of the form (\ref{eq:zero}), with $\omega = 0$
and $j = - \dot{\imath} \lambda \Lambda \xi_{0}
(\widehat{\mbox{\boldmath $\lambda$}})$.
In this case $\xi_{0}(\widehat{\mbox{\boldmath
$\lambda$}})$ is the symmetry mode
obtained by acting on the soliton fields with
the generator of isorotation about the
axis $\widehat{\mbox{\boldmath $\lambda$}}$:
\begin{equation}
\xi_{0}(\widehat{\mbox{\boldmath $\lambda$}}) =
\mbox{$1 \over 2$} \left ( \begin{array}{c}
  \dot{\imath} /2 \;\mbox{\boldmath $\tau$} \cdot
\widehat{\mbox{\boldmath $\lambda$}} q_h \\
  \dot{\imath} /2 \;\mbox{\boldmath $\tau$} \cdot
\widehat{\mbox{\boldmath $\lambda$}} q_h  \\
  - \widehat{\mbox{\boldmath $\lambda$}} \times
\mbox{\boldmath $\pi$}_h  \\
  0
                   \end{array} \right ) .
\label{eq:zeroisospin}
\end{equation}
Next, we have to find the conjugated mode,
by solving the second of Eqs. (\ref{eq:zero}).
We notice, that this is an odd-{\cal R} case (\ref{eq:oddR}).
We immediately
get $\mbox{\boldmath $P$} = \mbox{$1 \over 2$} \mbox{\boldmath $\lambda$}
\times \mbox{\boldmath $\pi$}_h$. For the quark shift, $\delta q_{cr}$,
a differential equation of the form (\ref{eq:oddR}) is solved \cite{CB86}.
The problem is of the type discussed in Sec. \ref{se:linearb}, where
${\cal M}$ is the moment of inertia, $\Theta$, and the ``charges'' are:
$\mbox{\boldmath $Q$}_1 = \mbox{\boldmath $\lambda$} \Theta$,
$Q_{\mu} = 0$ for $\mu \neq 1$.
The second-order energy shift is:
$\kappa = \mbox{$1\over 2$} \lambda^2 \Theta$.
Explicitly, one finds
\begin{eqnarray}
\Theta      &=&\Theta _{m} + \Theta _{q} , \nonumber \\
\Theta _{m} &=& \int d^3x (\widehat{\mbox{\boldmath
$\lambda$}} \times \mbox{\boldmath $\pi$}_h)^2 =
 (8 \pi /3) \int dr r^2 \pi_h^2 , \nonumber \\
\Theta _{q} &=& 2 \int d^3x
\delta q^{\dagger}_{cr} \widehat{\mbox{\boldmath
$\lambda$}} \cdot \mbox{\boldmath $\tau$} q_h
\label{eq:theta}
\end{eqnarray}

\subsection{Quantization}
\label{se:quant}

The simplest approach to quantization via cranking, is to recognize
that in the frame isorotating with velocity
$\mbox{\boldmath $\lambda$}$, in which we solve the cranking equations
of motion(Sec. \ref{se:crank}), we still have the freedom of
(iso)rotating the soliton by an arbitrary (time-independent)
angle. This is an example of the freedom
of choice in the $c_0$ coefficient in Sec. \ref{se:linearb},
which in this case corresponds to
three Euler angles, or, in the commonly used Cayley-Klein
notation \cite{ANW83}, to the matrix $B = b_0 + \mbox{\boldmath $b$}
\cdot \mbox{\boldmath $\tau$}$ \cite{CB86}.
In our mean-field approach, the corresponding fields carry
these (time-independent) $B$ matrices, and in the rotating
frame they assume the form:
\begin{equation}
 \sigma  \rightarrow  \sigma, \;\;\;
 \mbox{\boldmath $\pi$}  \rightarrow  B
\mbox{\boldmath $\pi$} B^{\dagger}, \;\;\;
 q \rightarrow B q .
\label{eq:angle}
\end{equation}
Matrix $B$ plays the role of coordinate variables conjugated to $\lambda$,
which upon quantization becomes a differential operator \cite{ANW83,CB86}.
The quantization is straightforwardly implemented in two steps:
1) one identifies the collective
spin and isospin operators, as done in Ref. \cite{CB86}. Then
\begin{equation}
\mbox{\boldmath $\lambda$}  \Theta = \mbox{\boldmath $J$},
\;\;\; I_a = c_{ab} J_b ,
\label{eq:JI}
\end{equation}
where
$J$ and $I$ are the spin and isospin operators, satisfying
appropriate commutation relations, and $c_{ab}$,
defined in App. \ref{ap:collective}, has the meaning of the
transformation matrix from the body-fixed to the lab frame \cite{Ring}.
2) Corresponding collective wave functions
are introduced. Expectation values of operators
are calculated by first identifying in the semiclassical
expression for an operator
its collective part (dependent on
$\mbox{\boldmath $\lambda$}$, $c_{ab}$, etc.),
and an intrinsic part (dependent on the meson and
quark fields $\sigma$, $\mbox{\boldmath $\pi$}$, $q$).
Then, the matrix element factorizes into a collective
matrix element in the wave functions of App. \ref{ap:collective} (this is
an integral over the collective coordinates, viz. Euler angles, or
$(b_0, \mbox{\boldmath $b$})$), and
an intrinsic matrix element, which is a space
integral over the quark and meson
fields. For details, see Ref. \cite{CB86}.

\subsection{External perturbations}
\label{se:external}

The quark and meson field profiles in Eq. (\ref{eq:angle}) are
in general not equal to the hedgehog profiles. We have demonstrated in
Sec. \ref{se:crank} that the quarks develop
shifts upon cranking. If some
other (external) interaction is present, then the profiles are
additionally shifted. These shifts are obtained
by solving the linear response
equations, as described in Sec. \ref{se:linear}.
We introduce a resolvent for the ${\cal H} - \omega \Lambda$
operator in Eq. (\ref{eq:linear}) (RPA propagator) and solve formally
Eq. (\ref{eq:linear}), obtaining
\begin{equation}
\xi = {\cal G} j , \;\;\; {\cal G} = ({\cal H} - \omega \Lambda)^{-1} .
\label{eq:formal}
\end{equation}
In the presence of cranking and some other external perturbation, we have
\begin{equation}
\xi = \xi_{cr} + \xi_{ext} = {\cal G} (j_{cr} + j_{ext}) ,
\label{eq:crext}
\end{equation}
where subscripts $cr$ and $ext$ refer to cranking,
and an external perturbation,
respectively.
The second-order energy shift corresponding to
a perturbation can be written as
\begin{equation}
\kappa = 2 \xi^{\dagger} j = 2 j^{\dagger} {\cal G} j .
\label{eq:kapj}
\end{equation}
The difference between this expression,
and the generic expression (\ref{eq:polarizability})
is that in the present case the source carries collective degrees of freedom,
$j = j^{coll} j^{intr}$. Thus, the matrix
element of $\kappa$ in a baryon state
$|b \rangle$ is (see example in Sec. \ref{se:ndeltasplit}):
\FL
\begin{eqnarray}
\kappa^b &=& 2 \langle coll|j^{coll}j^{coll}|coll \rangle \nonumber \\
& & \int d^3 x \; d^3x' \; j^{intr \dagger}(x)
{\cal G}(x,x') j^{intr}(x') ,
\nonumber \\
&=& 2 \langle coll|j^{coll}j^{coll}|coll \rangle \int d^3 x \;
\xi^{intr \dagger}(x) j^{intr}(x) ,
\label{eq:kapb}
\end{eqnarray}
where $|coll \rangle$ represents the collective
wave function (App. \ref{ap:collective}) associated
with the baryon state $|b \rangle$.

It is possible to have isospin-dependent effects in linear response of the
nucleon. For example, if the external
interaction has $K^{\cal P} = 1^{+}$ (the same
quantum numbers as in cranking), we
pick up cross terms between cranking, and the external perturbation
(see example in Sec. \ref{se:pn}):
\FL
\begin{eqnarray}
\kappa^b &=& 2 \langle coll|j^{coll}_{cr}
j^{coll}_{ext}|coll \rangle \nonumber \\
& & \int d^3 x \; d^3 x' \; j^{intr \dagger}_{cr}(x)
{\cal G}(x,x') j^{intr}_{ext}(x') +\; h.c. \nonumber \\
&=& 2 \langle coll|j^{coll}_{cr} j^{coll}_{ext}|coll \rangle
\int d^3 x \; \xi^{intr \dagger}_{cr}(x)
j^{intr}_{ext}(x) +\; h.c.  \nonumber \\
\label{eq:kapbcr}
\end{eqnarray}

Expressions (\ref{eq:kapb}, \ref{eq:kapbcr}) are just second-order
perturbation results. We may formally continue
to higher order in perturbation
theory, which leads to chains of the form
\FL
\begin{eqnarray}
\kappa_{{i_1}, ... ,{i_n}} &=&
 2 \langle coll|j_{i_1}^{coll} {\cal V}_{i_2}^{coll}
... j_{i_n}^{coll}|coll \rangle \nonumber \\
& & \int d^3 x_1 \; ... \; d^3 x_n \;
j_{i_1}^{intr \dagger}(x_1) {\cal G}(x_1,x_2)
{\cal V}_{i_2}(x_2) \nonumber \\
& & {\cal G}(x_2,x_3) \; ... \; {\cal V}_{i_{n-1}}(x_{n-1})
{\cal G}(x_{n-1},x_{n}) j_{i_n}^{intr}(x_n) , \nonumber \\
\label{eq:kappan}
\end{eqnarray}
where ${\cal V}_{i_k}$ is interaction of $k^{th}$
type. The total energy shift is
the sum over all possible orderings of
$({i_1}, ... ,{i_n})$ in (\ref{eq:kappan}).
Because the ground state has $K^{\cal P} = 0^{+}$,
the matrix element in
Eq. (\ref{eq:kappan}) is non-zero only if one can
compose the $K^{\cal P}$ quantum numbers
of $j_{i_1}, {\cal V}_{i_2}, ... , j_{i_n}$
to $K^{\cal P} = 0^{+}$.
In (I) we show an application of Eq. (\ref{eq:kappan}) with two RPA
propagators in the analysis of the neutron-proton splitting
of electromagnetic polarizabilities.
In Sec. \ref{se:nc}  we discuss in what cases going to a higher order
in perturbation theory is consistent with $N_c$-counting,
which is our basic principle
in organizing the perturbation expansion in hedgehog models.

\section{Simple examples}
\label{se:examples}

In this section we give some simple application of
the described formalism. A more advanced and physically important case
of electromagnetic polarizabilities is given in (I).

\subsection{$N$-$\Delta$ mass splitting}
\label{se:ndeltasplit}

As an illustration of application of Eq. (\ref{eq:kapb}), consider
the $N$-$\Delta$ mass splitting. In this case
$\kappa^b$ is the energy shift
of the baryon $|b \rangle $ due to the cranking
perturbation. From
Eqs. (\ref{eq:JI}, \ref{eq:theta}, \ref{eq:kapb})
we obtain immediately
the usual expression for the $N$-$\Delta$ mass splitting:
\begin{equation}
M_{\Delta} - M_N = \mbox{$1\over 2$}
(\langle \Delta |\lambda^2 |\Delta \rangle -
\langle N |\lambda^2 |N \rangle) \Theta = \frac{3}{2 \Theta}
\label{eq:ndeltamass}
\end{equation}

\subsection{Hadronic $p$ - $n$ mass splitting}
\label{se:pn}

As an example of an isospin-dependent effect,
consider the neutron-proton
mass difference due to the difference of
the up and down quark masses.
The perturbation in the lagrangian has the form
${\cal L}_m = \mbox{$1\over 2$}
(m_d - m_u) \overline{\psi} \tau_3 \psi$. It has
$K^{\cal PR} = 1^{+-}$, exactly as cranking,
hence a mixed perturbation of the form
(\ref{eq:kapbcr}) appears. Passing to an
isorotating frame, we find the source corresponding to
the quark mass splitting, which arises in Eqs. (\ref{eq:oddR}):
$j_m = \mbox{$1\over 2$} (m_d - m_u) N_c \gamma_0
\mbox{\boldmath $c$} \cdot \mbox{\boldmath $\tau$} q_h$,
where $\mbox{\boldmath $c$}$ is
defined in App. \ref{ap:collective}.
Since we have already solved the cranking equation,
we do not have to solve the new
equation with source $j_m$. We simply calculate
the overlap of $j_m$ with
the shift in the fields due to cranking, $\delta q_{cr}$,
according to
Eq. (\ref{eq:kapbcr}). Using the fact that
$\langle N |\mbox{\boldmath $\lambda$} \cdot
\mbox{\boldmath $c$}|N \rangle = -\langle N |I_3|N \rangle $,
we obtain the following expression for the hadronic
splitting of the neutron and proton masses:
\FL
\begin{eqnarray}
(M_n - M_p)^{hadr.} &=& (m_d - m_u)
\int d^3 x \; d^3 x' [ \langle n|j_m^{\dagger} \delta q_{cr}
|n \rangle \nonumber \\
& & - (n \leftrightarrow p) ] \nonumber \\
&=&  \frac{m_d - m_u}{\Theta}
\int d^3 x \; j_m^{intr \dagger} \delta q_{cr}^{intr} .
\label{eq:npmass}
\end{eqnarray}
The numerical value, obtained for the solution of Ref. \cite{CB86} gives
$(M_n - M_m)^{hadr.} = 0.4 \times (m_d - m_u)$, which for typical values
of $(m_d - m_u)$ gives a number around $2 MeV$.
The electromagnetic mass difference
can also be studied in hedgehog models models \cite{Goeke:pnEM}.

\section{$N_c$-counting}
\label{se:nc}

The basic organizational principle behind hedgehog models is
the $1/N_c$ expansion of QCD \cite{thoft:nc,witten2:nc,witten:nc}.
In the $N_c \rightarrow \infty$ limit, masses of baryons diverge as
$N_c$, and can be calculated using mean-field theory \cite{witten2:nc}.
It should be noted that the assumption of the spin-isospin
correlated wave function, which is essential in hedgehog models,
does not follow from the large-$N_c$ limit alone
--- it is an additional
assumption of the hedgehog approach. By analogy to nuclear physics,
in systems with many nucleons we may have nuclei
with intrinsic deformations,
but we may also have spherically symmetric nuclei,
and it is the dynamics
which determines whether the wave-function is deformed or not.
In hedgehog models the hedgehog wave function is assumed to be
deformed in the spin-isospin space, and the nucleon the $\Delta$ masses,
which are of the order $N_c$,  are
degenerate in the leading-$N_c$ order.

When cranking is used, these masses split as $\sim N_c^{-1}$. In fact,
cranking becomes an exact projection method in the large-$N_c$ limit,
since it may be viewed as a Peierls-Yoccoz
projection with $\delta$-function
overlaps between rotated wave functions \cite{Ring}. Thus we obtain the
hedgehog result for the mass splitting, Eq. (\ref{eq:ndeltamass}).
It would not be consistent, however, to conclude that the nucleon
or $\Delta$ masses individually are given by the hedgehog soliton mass plus
the cranking piece. There are other effects (center-of-mass correction,
centrifugal stretching, etc.) which enter at the same level as the
cranking term. Also, the effective lagrangian
may be supplemented by subleading
terms in $N_c$, which we did not have to
include to obtain the leading piece
in the hedgehog mass. Therefore, it is useless to write down
${M_J = M_h + J(J+1)/(2 \Theta) + {\cal O}(N_c^{-1})}$, since the
last term, which we do not calculate, enters at the same level as
the cranking term.  We can only trust the leading piece,
${M_J = M_h + {\cal O}(N_c^{-1})}$,
and, in order to maintain consistency with the $N_c$-counting,
the mass formula should not be ``improved'' by adding the cranking term.
The mentioned effects of center-of-mass corrections, centrifugal
stretching, etc., are at the leading level the same for the nucleon
and for the $\Delta$, therefore for the $N-\Delta$
mass splitting we get the formula
${M_{\Delta} - M_N = 3/(2 \Theta) + {\cal O}(N_c^{-2})}$.

The prescription, which we tried to illustrate above, is that
with semiclassical methods we can only get the leading-$N_c$
term for a given observable. The power of $N_c$ varies, depending
on the quantity we are investigating.
The same is true for the calculation
of polarizabilities, described in this paper. We can easily
obtain the $N_c$ behavior of various terms in
Eqs. (\ref{eq:kapb},\ref{eq:kapbcr},\ref{eq:kappan}),
but only the leading-$N_c$ piece corresponding to a particular polarizability
should be retained. As an illustration, consider the electric polarizability
of the nucleon, discussed extensively in \cite{BBC91nt} and in (I).
The electric field polarizes the hedgehog. The
electric charge of the quark has an isoscalar component, of order $N_c^{-1}$,
and isovector component, of order $1$. We immediately see from
Eq. (\ref{eq:kapb}) that the leading part of the electric polarizability
of the nucleon is obtained from interactions with two isovector sources,
and the term with two isoscalar sources is two powers of $N_c$ suppressed.
The non-dispersive seagull effects also enter at the level of $N_c$ (I),
hence the nucleon polarizability goes as $N_c$. Quite analogously
to the problem of the $N-\Delta$ mass splitting,
the neutron-proton splitting of the electric polarizability is a $N_c^{-1}$
effect (I), and we can calculate it consistently only to this order.

In principle, one might try to perform a calculation which
consistently takes into account the subleading pieces. The
appropriate scheme would be the Kerman-Klein method \cite{KermanKlein},
but its application would involve a
complicated fully quantum-mechanical calculation.

\section{Other models}
\label{se:other}

Techniques described in this paper are applicable to other model
after straightforward modifications. In the Skyrme model, the
described RPA method involves fluctuations of the
meson fields which do not satisfy the nonlinear constraint for the
$\sigma$ and $\pi$ field operators. This linearization may be viewed
as an approximation to the fully nonlinear dynamics.
The RPA dynamics,  obviously, involves mesons only, and the higher-derivative
terms are manifest in the equations of motion for the fluctuations.

In the case of the (partly bosonized) NJL model \cite{Eguchi:boson},
the mesonic potential has the simple form
$\mbox{$1 \over 2$} \mu^2 (\sigma^2 + \mbox{\boldmath $\pi$}^2)$.
The sea quarks are present explicitly, and the number of quark equations
is infinite. Standard methods of solving these equations numerically may
encounter problems for the case when the
translational zero mode is excited, since
extremely good accuracy is necessary in this case.

\section{Conclusion}
\label{se:conclusion}

We have presented the linear response method in hedgehog soliton
models. We have shown that the method is consistent with the
basic philosophy of these models, namely, the $1/N_c$-expansion,
if its application is restricted to obtaining the
leading-$N_c$ order of a given
quantity. We have discussed many technical points which are encountered in
practical calculations, especially the treatment
of zero-modes, which create special problems. Appropriate
equations of motion have been classified
according to hedgehog symmetries, and derived explicitly for the model of
Ref. \cite{BirBan8485}. Our method, after straightforward modifications, is
directly applicable to other hedgehog models.
A physical application of the approach is described in the
preceding paper, (I), where we study the
electromagnetic polarizabilities of the nucleon.

\acknowledgements

Support of the the National Science Foundation (Presidential Young
Investigator grant), and of the U.S. Department of Energy is gratefully
acknowledged. We thank Manoj Banerjee for many useful
suggestions and countless valuable comments.
One of us (WB) acknowledges a partial support of
the Polish State Committee for Scientific Research
(grants 2.0204.91.01 and 2.0091.91.01).

\newpage

\appendix{Equations of motion for small fluctuations in the grand-spin basis}
\label{ap:eqmotion}

We compose the basis of Dirac spinors with good $K$ quantum numbers
using the coupling scheme in which the isospin,
$I=\mbox{$1\over 2$}$, and
spin $S=\mbox{$1\over 2$}$,
are first coupled to a quantum number $\Lambda$, and then orbital
angular momentum, $L$, and $\Lambda$ are coupled to $K$.
Since there is no confusion concerning the value of $K$ or $K_3$,
we use the notation
\begin{equation}
|L,\Lambda > = |K,(L,\Lambda (I=\mbox{$1\over 2$},
S=\mbox{$1\over 2$})),K_{3}> .
\label{eq:ap:coupling}
\end{equation}
States with parity ${\cal P} = (-)^K$ (or ${\cal P} = -(-)^K$)
are called normal (abnormal) parity states.
The basis of Dirac spinors is
\begin{equation}
q^{L,\Lambda} = \left ( \begin{array}{c} G^{L,\Lambda}(r) \\
       \dot{\imath} \mbox{\boldmath $\sigma$}
\cdot \widehat{\mbox{\boldmath $r$}} F^{L,\Lambda}(r)
           \end{array} \right ) |L,\Lambda > .
\label{eq:ap:spinors}
\end{equation}
Spinors $X$ and $Y$ are expressed in states (\ref{eq:ap:spinors}).
The quark sources are decomposed into $(L,\Lambda)$ components:
\begin{equation}
j^{L,\Lambda} = \left ( \begin{array}{c} j_G^{L,\Lambda}(r) \\
       \dot{\imath} \mbox{\boldmath $\sigma$}
\cdot \widehat{\mbox{\boldmath $r$}} j_F^{L,\Lambda}(r)
           \end{array} \right ) |L,\Lambda > ,
\label{eq:ap:spinorsource}
\end{equation}
Tables \ref{ta:taudotr} - \ref{ta:sigmadotL} list the matrix elements
which arise in deriving the quark parts of perturbation equations.
It is clear from Table \ref{ta:sigmadotL}
that unless $K=0$, the kinetic term
mixes the $\Lambda = 0$ and $\Lambda = 1$ components of the $L=K$
states (normal parity case). Diagonalization is made through the
substitution
\begin{eqnarray}
G^{a} = \sqrt{\frac{K+1}{2 K+1}} G^{K,0} -
 \sqrt{\frac{K}{2 K+1}} G^{K,1} , \nonumber \\
G^{b} = \sqrt{\frac{K}{2 K+1}} G^{K,0} +
 \sqrt{\frac{K+1}{2 K+1}} G^{K,1} ,
\label{eq:ap:diagonal}
\end{eqnarray}
and similarly for the $F$-components, and the sources.

The basis for the meson fluctuations is composed
by coupling isospin to $L$. For a given value of K,
the $\sigma$ and $\pi$ fluctuations can be expressed through
functions
\begin{equation}
\sigma^L (r) |K,(L,0),K_{3}>, \pi^L (r) |K,(L,1),K_{3}> ,
\label{eq:ap:mesons}
\end{equation}
Obviously, $L=K$ for $\sigma$, and $L=K-1,K,K+1$ for
$\pi$, such that for a given $K$ and $K_3$
\begin{eqnarray}
\delta \sigma &=& \sigma^K (r) |K,(L,0),K_{3}>, \nonumber \\
\delta \pi &=&\/\/\/ \sum_{L=K-1,K,K+1}\/\/\/ \pi^L (r) |K,(L,1),K_{3}> .
\label{eq:ap:mesfl}
\end{eqnarray}
Using standard Racah algebra, it is straightforward
to derive the general equations
(\ref{eq:linear}) for a given $K$ perturbation.
In the notation of this appendix,
$G^{a}_{\{X,Y\}}$, {etc.}, correspond to the $X$ and $Y$
spinors from Eq. (\ref{eq:linear}) ,
and $G^{a}_{\{X+Y\}} = G^{a}_X + G^{a}_Y$, {etc.}. The
functions describing meson fluctuation,
$\sigma^L, \pi^L$, have the meaning of the
$Z$-functions of Eq. (\ref{eq:linear}).

For normal parity equation we get
\mediumtext
\begin{eqnarray}
\partial_r G^{a}_{\{X,Y\}} &=& \frac{K}{r} G^{a}_{\{X,Y\}} +
  (g \sigma_h - \varepsilon \mp \omega) F^{a}_{\{X,Y\}} \nonumber \\
  &+& g \pi_h ( - \frac{1}{2K+1} G^{a}_{\{X,Y\}}
           - \frac{2 \sqrt{K(K+1)}}{2K+1} G^{b}_{\{X,Y\}} ) \nonumber \\
&+& g ( \sqrt{\frac{K+1}{2 K+1}} F_h \sigma^K + G_h \pi^{K+1} )
   - j^{a}_{F,\{ X,Y \} } , \nonumber \\
\partial_r G^{b}_{\{X,Y\}} &=& -\frac{K+1}{r} G^{b}_{\{X,Y\}} +
  (g \sigma_h - \varepsilon \mp \omega) F^{b}_{\{X,Y\}} \nonumber \\
  &+& g \pi_h ( - \frac{2 \sqrt{K(K+1)}}{2K+1} G^{a}_{\{X,Y\}}
           + \frac{1}{2K+1} G^{b}_{\{X,Y\}} ) \nonumber \\
&+& g ( \sqrt{\frac{K}{2 K+1}} F_h \sigma^K - G_h \pi^{K-1} )
   - j^{b}_{F,\{ X,Y \} } , \nonumber \\
\partial_r F^{a}_{\{X,Y\}} &=& - \frac{K+2}{r} F^{a}_{\{X,Y\}} +
  (g \sigma_h + \varepsilon \pm \omega) G^{a}_{\{X,Y\}} \nonumber \\
  &+& g \pi_h (  \frac{1}{2K+1} F^{a}_{\{X,Y\}}
           + \frac{2 \sqrt{K(K+1)}}{2K+1} F^{b}_{\{X,Y\}} ) \nonumber \\
&+& g ( \sqrt{\frac{K+1}{2 K+1}} G_h \sigma^K - F_h \pi^{K+1} )
   + j^{a}_{G,\{ X,Y \} } , \nonumber \\
\partial_r F^{b}_{\{X,Y\}} &=& \frac{K-1}{r} F^{b}_{\{X,Y\}} +
  (g \sigma_h + \varepsilon \pm \omega) G^{b}_{\{X,Y\}} \nonumber \\
   &+& g \pi_h ( \frac{2 \sqrt{K(K+1)}}{2K+1} F^{a}_{\{X,Y\}}
           - \frac{1}{2K+1} F^{b}_{\{X,Y\}} ) \nonumber \\
&+& g ( \sqrt{\frac{K}{2 K+1}} G_h \sigma^K + F_h \pi^{K-1} )
   + j^{b}_{G,\{ X,Y \} } , \nonumber \\
\label{eq:ap:normalparityq}
\end{eqnarray}
\begin{eqnarray}
\left ( \partial_r^2 \right. &+& \left. \frac{2}{r} \partial_r -
 \frac{(K-1)K}{r^2} \right ) \pi^{K-1} =
\lambda^2 (\sigma_h^2 + \pi_h^2 - \nu^2 - \omega^2) \pi^{K-1} \nonumber \\
&+& 2 \lambda^2 \left (
   \frac{K^2}{2K+1} \pi_h^2 \pi^{K-1}
   + \sqrt{\frac{K}{2K+1}} \sigma_h \pi_h \sigma^K
   - \frac{\sqrt{K(K+1)}}{2K+1} \pi_h^2 \pi^{K+1} \right ) \nonumber \\
&-& g N_c ( F_h G^{b}_{\{X+Y\}} + G_h F^{b}_{\{X+Y\}}  )
    + j^{K-1}_{\pi} , \nonumber \\
 \nonumber \\
\left ( \partial_r^2 \right. &+& \left. \frac{2}{r} \partial_r -
\frac{K(K+1)}{r^2} \right ) \sigma^K =
\lambda^2 (\sigma_h^2 + \pi_h^2 - \nu^2 - \omega^2) \sigma^K \nonumber \\
&+& 2 \lambda^2 \left (
  \sqrt{\frac{K}{2K+1}} \sigma_h \pi_h \pi^{K-1}
   +  \sigma_h^2 \sigma^K
   - \sqrt{\frac{K+1}{2K+1}} \sigma_h \pi_h \pi^{K+1}\right ) \nonumber \\
&-& g N_c \left (
      G_h \left ( \sqrt{\frac{K+1}{2K+1}} G^{a}_{\{X+Y\}} +
              \sqrt{\frac{K}{2K+1}}   G^{b}_{\{X+Y\}}  \right ) \right.
\nonumber \\
& &-  \left. F_h \left ( \sqrt{\frac{K+1}{2K+1}} F^{a}_{\{X+Y\}} +
              \sqrt{\frac{K}{2K+1}}   F^{b}_{\{X+Y\}}  \right ) \right )
    + j^{K}_{\sigma} , \nonumber \\
 \nonumber \\
\left ( \partial_r^2 \right. &+& \left. \frac{2}{r} \partial_r -
 \frac{(K+1)(K+2}{r^2} \right ) \pi^{K+1} =
\lambda^2 (\sigma_h^2 + \pi_h^2 - \nu^2 - \omega^2) \pi^{K+1} \nonumber \\
&+& 2 \lambda^2 \left (
   - \frac{\sqrt{K(K+1)}}{2K+1} \pi_h^2 \pi^{K-1}
   - \sqrt{\frac{K+1}{2K+1}} \sigma_h \pi_h \sigma^K
   - \frac{K+1}{2K+1} \pi_h^2 \pi^{K+1} \right )  \nonumber \\
&-& g N_c ( -F_h G^{a}_{\{X+Y\}} - G_h F^{a}_{\{X+Y\}}  )
    + j^{K+1}_{\pi} , \nonumber \\
\label{eq:ap:normalparitym}
\end{eqnarray}

The abnormal parity equations have the form
\begin{eqnarray}
\partial_r G^{K-1,1}_{\{X,Y\}} &=& \frac{K-1}{r} G^{K-1,1}_{\{X,Y\}} +
  (g \sigma_h - \varepsilon \mp \omega) F^{K-1,1}_{\{X,Y\}} \nonumber \\
  &+& g \pi_h ( \frac{1}{2K+1} G^{K-1,1}_{\{X,Y\}}
           + \frac{2 \sqrt{K(K+1)}}{2K+1} G^{K+1,1}_{\{X,Y\}} ) \nonumber \\
&-& g  \sqrt{\frac{K+1}{2 K+1}} G_h \pi^K
   - j^{K-1,1}_{F,\{ X,Y \} } , \nonumber \\
\partial_r G^{K+1,1}_{\{X,Y\}} &=& -\frac{K+2}{r} G^{K+1,1}_{\{X,Y\}} +
  (g \sigma_h - \varepsilon \mp \omega) F^{K+1,1}_{\{X,Y\}} \nonumber \\
  &+& g \pi_h (  \frac{2 \sqrt{K(K+1)}}{2K+1} G^{K-1,1}_{\{X,Y\}}
           - \frac{1}{2K+1} G^{K+1,1}_{\{X,Y\}} ) \nonumber \\
&-& g \sqrt{\frac{K}{2 K+1}} G_h \pi^K
   - j^{K+1,1}_{F,\{ X,Y \} } , \nonumber \\
\partial_r F^{K-1,1}_{\{X,Y\}} &=& -\frac{K+1}{r} F^{K-1,1}_{\{X,Y\}} +
  (g \sigma_h + \varepsilon \pm \omega) G^{K-1,1}_{\{X,Y\}} \nonumber \\
  &+& g \pi_h ( - \frac{1}{2K+1} F^{K-1,1}_{\{X,Y\}}
           - \frac{2 \sqrt{K(K+1)}}{2K+1} F^{K+1,1}_{\{X,Y\}} ) \nonumber \\
&+& g  \sqrt{\frac{K+1}{2 K+1}} F_h \pi^K
   + j^{K-1,1}_{G,\{ X,Y \} } , \nonumber \\
\partial_r F^{K+1,1}_{\{X,Y\}} &=& \frac{K}{r} F^{K+1,1}_{\{X,Y\}} +
  (g \sigma_h + \varepsilon \pm \omega) G^{K+1,1}_{\{X,Y\}} \nonumber \\
  &+& g \pi_h (  - \frac{2 \sqrt{K(K+1)}}{2K+1} F^{K-1,1}_{\{X,Y\}}
           + \frac{1}{2K+1} F^{K+1,1}_{\{X,Y\}} ) \nonumber \\
&+& g \sqrt{\frac{K}{2 K+1}} F_h \pi^K
   + j^{K+1,1}_{G,\{ X,Y \} } ,
\label{eq:ap:abnormalparityq}
\end{eqnarray}
\FL
\begin{eqnarray}
\left ( \partial_r^2 \right. &+& \left. \frac{2}{r} \partial_r -
 \frac{K(K+1)}{r^2} \right ) \pi^{K} =
\lambda^2 (\sigma_h^2 + \pi_h^2 - \nu^2 - \omega^2) \pi^{K} \nonumber \\
&-& g N_c \left (
      G_h \left ( \sqrt{\frac{K+1}{2K+1}} F^{K-1,1}_{\{X+Y\}} +
              \sqrt{\frac{K}{2K+1}}     F^{K+1,1}_{\{X+Y\}}  \right )
\right. \nonumber \\
   &+& \left.  F_h \left ( \sqrt{\frac{K+1}{2K+1}} G^{K-1,1}_{\{X+Y\}} +
              \sqrt{\frac{K}{2K+1}}
G^{K+1,1}_{\{X+Y\}}  \right ) \right )
    + j^{K}_{\pi} . \nonumber \\
\label{eq:ap:abnormalparitym}
\end{eqnarray}
\narrowtext

For the $\omega = 0$, odd-${\cal R}$ case,
meson fluctuations vanish, and
appropriate equations have the form of
Eqs. (\ref{eq:ap:normalparityq},\ref{eq:ap:abnormalparityq}), with
the meson fluctuations set to zero. The $X$ and $Y$
equations can be combined to
a single equation of the form (\ref{eq:oddR}).

In the case of an even-${\cal R}$ source which does
not excite a zero mode (case
$Q_0 = 0$ in Sec. \ref{se:lineara}), we can set
$\omega = 0$, in the above equations. We can combine
the $X$ and $Y$ equations, and obtain the
form (\ref{eq:evenR}). If the zero mode is excited ($Q_0 \neq 0$),
we have to solve full equations
(\ref{eq:ap:normalparityq}-\ref{eq:ap:normalparitym}), or
(\ref{eq:ap:abnormalparityq}-\ref{eq:ap:abnormalparitym}),
depending on parity.

For the special case of $K=0$, $G^a = G^{0,0}$, $G^b=0$,
etc., and only equations for
the $a$ components in Eqs. (\ref{eq:ap:normalparityq})
remain. Fields with negative (i.e. $K-1$)
superscripts, and equations for these fields are eliminated.

The boundary conditions in
Eqs. (\ref{eq:ap:normalparityq}-\ref{eq:ap:abnormalparitym})
are such that the solutions are everywhere finite. At the origin,
radial derivatives of $S$-wave fields vanish, and
the values of higher-L fields vanish. At
$r \rightarrow \infty$, the appropriate
boundary conditions follow from solutions
of the equations in the asymptotic region.

\appendix{Collective matrix elements}
\label{ap:collective}

Suppose a space rotation, $R$, is described by
Euler angles $\alpha$, $\beta$ and $\gamma$:
\begin{equation}
R = e^{-\dot{\imath} \alpha J_z}
e^{-\dot{\imath} \beta J_y} e^{-\dot{\imath} \gamma J_z} .
\label{eq:ap:R}
\end{equation}
Then, matrix $B$ from Sec. \ref{se:quant} is given by
\begin{equation}
B = e^{\dot{\imath} \gamma \tau_3 /2} e^{\dot{\imath}
\beta \tau_2 /2 } e^{\dot{\imath} \alpha \tau_3 /2} .
\label{eq:ap:B}
\end{equation}
The matrix transforming from the body-fixed to
lab frame, $c_{ab}$, is defined as
\begin{equation}
c_{ab} = \mbox{$1\over 2$} Tr[\tau_a B
\tau_b B^{\dagger}] = D^{1}_{ba}(\alpha, \beta, \gamma) ,
\label{eq:ap:cab}
\end{equation}
where the first (second) subscript in the
Wigner D-matrix is connected to the spin (isospin) space.
It follows that $\sum_{b} c_{ab} J_b = - I_a$. The spin operator and
the matrix $c$ commute, $[c_{ab},J_k] = 0$.
We also introduce a vector $\mbox{\boldmath $c$}$ defined as
\begin{equation}
\mbox{\boldmath $c$} = \mbox{$1\over 2$}
Tr[\tau_3 B \mbox{\boldmath $\tau$} B^{\dagger}]; \;\;\;
c_{\mu} = D^{1}_{\mu 0} .
\label{eq:ap:c}
\end{equation}
The collective baryon states with spin $J$,
isospin $I=J$, and projections $m$ and $I_3$ are
\begin{equation}
|J=I;m,I_3 \rangle = \sqrt{\frac{2J+1}{8 \pi^2}} D^J_{m,-I_3} .
\label{eq:ap:state}
\end{equation}
In formulas below we do not display $m$ or $I_3$
in labels of the states, and use notation
$|N \rangle = |\mbox{$1\over 2$}; m, I_3 \rangle$,
and $|\Delta \rangle = |{\mbox{$3\over 2$}}; m, I_3 \rangle$.
The following useful formulas can be easily derived
(no implicit summation over repeated indices):
\FL
\begin{equation}
(c_{\mu})^{*} c_{\mu} = {\mbox{$1\over 3$}} +
({\mbox{$2\over 3$}} - \mu^2) D^2_{00} ,
\;\;\; \sum_{\mu} (c_{\mu})^{*} c_{\mu} = 1 ,
\label{eq:ap:cc1}
\end{equation}
from which follows that
\FL
\begin{eqnarray}
\langle N|(c_{\mu})^{*} c_{\mu}|N \rangle &=&
{\mbox{$1\over 3$}} ; \;\;\; {\rm any} \;\;\mu , \nonumber \\
\langle \Delta|(c_{\mu})^{*} c_{\mu}|\Delta \rangle &=&
  {\mbox{$1\over 3$}} + \frac{{\mbox{$2\over 3$}} -
\mu^2}{5} \left\{ \begin{array}{l} + 1 \;{\rm ;}\;\; |m|=|I_{3}| \\
         - 1 \;{\rm ;} \;\;|m| \neq |I_{3}|
    \end{array} \right. , \nonumber \\
\langle N|(c_{\mu})^{*} c_{\mu}|\Delta \rangle &=&
  \frac{\sqrt{2}({\mbox{$2\over 3$}} -
\mu^2)}{5} \left\{ \begin{array}{l} +
1 \;{\rm ;}\;\; m=I_{3} \\
              - 1 \;{\rm ;} \;\;m = -I_{3}
 \end{array} \right. .
\label{eq:ap:cc2}
\end{eqnarray}
One also finds
\begin{equation}
\langle N|\left ( (J_{\mu})^{*} c_{\mu} +
(c_{\mu})^{*} J_{\mu} \right )|N \rangle =
- {\mbox{$2\over 3$}} I_3 ;
\;\;\; {\rm any} \;\;\mu .
\label{eq:ap:Jc}
\end{equation}
One also derives
\begin{equation}
\langle N| c_{0}|\Delta \rangle = \frac{\sqrt{2}}{3} .
\label{eq:ap:c0}
\end{equation}
For our analysis of the $\Delta$ states in hadronic
loops in (I) the following formulas are important:
\FL
\begin{eqnarray}
\sum_{\mu , m', I_{3}'} \langle N|(c_{\mu})^{*}|N; m', I_{3}' \rangle
                         \langle N; m', I_{3}'|c_{\mu}|N \rangle
&=& {\mbox{$1\over 3$}} , \nonumber \\
\sum_{\mu , m', I_{3}'} \langle N|(c_{\mu})^{*}|\Delta; m',
I_{3}' \rangle
                         \langle \Delta; m', I_{3}'|c_{\mu}|N
\rangle &=& {\mbox{$2\over 3$}} .
\label{eq:ap:intermediate}
\end{eqnarray}
The sum of the above formulas gives unity,
in accordance to the sum rule (\ref{eq:ap:cc1}).

\appendix{Equality of inertial and soliton masses}
\label{ap:masses}

For the case of translations, the inertia parameter,
${\cal M}$, is equal to the soliton mass, $M_{sol}$. This result,
required by Lorentz invariance, can be verified explicitly as follows:
Consider a boost in the z-direction, with small velocity $v$.
The fields transform as
\begin{eqnarray}
\phi &\rightarrow& \phi_h(\mbox{\boldmath $r$} - v t) , \nonumber \\
e^{- \dot{\imath} \varepsilon t}
&\rightarrow& e^{- \dot{\imath} \varepsilon (t - v z)
+ \mbox{$1\over 2$} v \alpha_{z}}
q_h(\mbox{\boldmath $r$} - v t) ,
\label{eq:ap:boost}
\end{eqnarray}
which lead to the following shifts linear in the velocity:
\begin{eqnarray}
\delta \phi &=& - v t \partial_z \phi_h , \nonumber \\
\delta q_h  &=& v e^{- \dot{\imath} \varepsilon t}
(\dot{\imath} \varepsilon z + \mbox{$1\over 2$}
\alpha_z - t \partial_z) q_h .
\label{eq:ap:boostshift}
\end{eqnarray}
Using identities $[ h, z ] = - \dot{\imath} \alpha_z$ and
$\{ h,\alpha_z \} = - 2 \dot{\imath}
\partial_z$, we easily derive the equation
\begin{equation}
[ h - \varepsilon] (- \varepsilon z +
\mbox{$1\over 2$} \dot{\imath} \alpha_{z}) q_h
 = \partial_z q_h .
\label{eq:ap:boosteq}
\end{equation}
After integrating by parts we get
the expression for the energy shift of a moving soliton:
\begin{equation}
\delta {\cal E} = \mbox{$1\over 2$} v^2 ( {\mbox{$1\over 3$}} T_q +
{\mbox{$2\over 3$}} T_{\phi}
 + N_c \varepsilon) ,
\label{eq:ap:solenergy}
\end{equation}
where $T_q$ and $T_{\phi}$ are kinetic energies in the soliton,
carried by the quarks and mesons, respectively.

Next, we use a virial relation. Consider scale change of the radial
coordinate, $r \rightarrow s r$. The soliton energy scales as
${\cal E}(s) = T_q / s + V_{q \phi} + s T_{\phi} + s^3 V_{\phi}$, where
$V_{q \phi}$ and $V_{\phi}$ are the quark-meson, and meson-meson interaction
energies. Stationarity of the solution imposes
$\partial_s {\cal E} |_{s=1} = 0$, which, together with the relation
$N_c \varepsilon = T_q + V_{q \phi}$, leads to the virial relation
\begin{equation}
M_{sol} = {\mbox{$1\over 3$}} T_q + {\mbox{$2\over 3$}} T_{\phi} + N_c
\varepsilon .
\label{eq:ap:virial}
\end{equation}
Comparing Eq. (\ref{eq:ap:solenergy}) and Eq. (\ref{eq:ap:virial})
completes the proof that $\delta {\cal E} = \mbox{$1\over 2$} v^2 M_{sol}$.
Using similar methods, one can show the equality of inertial and
soliton masses in other models \cite{Jackiw:rev},
also in non-local theories, such as the NJL model \cite{Pobnt}.

\appendix{Bosonization and Pauli blocking of the Dirac sea}
\label{ap:Pauli}

In this section we return to the question whether the Dirac sea
should be ``Pauli-blocked'' in our model.  Effective chiral models
are believed to result from bosonizing QCD, which,
of course, can only be done approximately. For definiteness, we discuss
the issue of Pauli blocking in the framework of
the partly-bosonized \cite{Eguchi:boson}
NJL model, but the result is more general.
In presence of an external source, $J$, the action of the model is:
\begin{eqnarray}
S_{NJL} &=& - \dot{\imath} Tr \,
log[ \dot{\imath} \overlay{\slash}{\partial} - g U - J]
- {vac} , \nonumber \\
g U &=& g(\sigma + \dot{\imath} \gamma_5
\mbox{\boldmath $\tau$} \cdot \mbox{\boldmath $\pi$}) ,
\label{eq:actionNJL}
\end{eqnarray}
A cut-off is understood, $Tr$ denotes functional trace, and $vac$ means
the vacuum subtraction. For simplicity, we
assume the nonlinear constraint
\mbox{$\sigma^2 + \mbox{\boldmath $\pi$}^2 = {F_\pi}^2$}. The
source ${J}$ may represent interactions with external probes ({\sl e.g.}
electromagnetic) or result from cranking (Sec. \ref{se:crank}).
For definiteness, let us evaluate the
moment of inertia. In this case
${J} = \mbox{$1\over 2$} \mbox{\boldmath $\lambda$}
\cdot \mbox{\boldmath $\tau$}$, and expanding
the action to second order in $\lambda$ we obtain
\begin{equation}
\Delta S_{NJL} = \mbox{$1\over 2$} \lambda^2 \Theta \int dt ,
\label{eq:expandedNJL}
\end{equation}
where the moment of inertia, $\Theta$, is given by
\begin{equation}
\Theta = \frac{\dot{\imath}}{4} N_c \int \frac{d \omega}{2 \pi}
Sp \frac{1}{\omega - h} \tau_3 \frac{1}{\omega - h} \tau_3 ,
\label{eq:thetaNJL}
\end{equation}
where $Sp$ denote the trace over space, spin and isospin, and $h$ is
the Dirac hamiltonian. The pole structure and
the contour of the $\omega$ integration in Eq. (\ref{eq:thetaNJL})
is given in Fig. (\ref{fi:contval}). Note that the contour goes
above the occupied valence state, as well as above all the negative-energy
sea states. Performing the integration over
$\omega$ in Eq. (\ref{eq:thetaNJL}),
we obtain the usual spectral expression for:
\begin{equation}
\Theta = \mbox{$1\over 2$} N_c \sum_{\begin{array}{c}
\scriptstyle i \in  {\rm occ.} \\
   \scriptstyle j \in  {\rm unocc.}
                           \end{array} }
 \frac{{\left | \langle i | \tau_3 | j \rangle \right |}^2}
 {\varepsilon_i - \varepsilon_j} ,
\label{eq:thetaspectral}
\end{equation}
where $occ.$ denotes all occupied states,
{\sl i.e.} the valence as well as
the sea states, and $unocc.$ denotes the
unoccupied positive energy states
(see Fig. \ref{fi:contval} for the meaning of labels).
The expression under the sum is antisymmetric with respect to
exchanging $i$ and $j$, therefore the sum as
in Eq. (\ref{eq:thetaspectral})
over $i$ and $j$ belonging to the same set
of indices vanishes. Using
this trick we can replace the ranges of
summation indices as follows:
\FL
\begin{equation}
  \sum_{\begin{array}{c} \scriptstyle i \in  {\rm occ.} \\
                       \scriptstyle j \in  {\rm unocc.} \end{array} }
=
  \sum_{\begin{array}{c} \scriptstyle i \in  {\rm occ.} \\
                       \scriptstyle j \in  {\rm all}
\end{array} }\!\!\! {}^{'} \,
=
  \sum_{\begin{array}{c} \scriptstyle i \in  {\rm val.} \\
                         \scriptstyle j \in  {\rm all}
\end{array} }\!\!\! {}^{'} \,
+ \sum_{\begin{array}{c} \scriptstyle i \in  {\rm sea} \\
                         \scriptstyle j \in  {\rm all}
\end{array} }\!\!\! {}^{'} \,
=
  \sum_{\begin{array}{c} \scriptstyle i \in  {\rm val.} \\
                         \scriptstyle j \in  {\rm all}
\end{array} }\!\!\! {}^{'} \,
+ \sum_{\begin{array}{c} \scriptstyle i \in  {\rm sea} \\
                         \scriptstyle j \in  {\rm pos. en.}
\end{array} } ,
\label{eq:trick}
\end{equation}
where the prime means the exclusion of $i=j$ term,
$any$ denotes all states, and
$pos. en.$ denotes the positive energy states. According to
Eq. (\ref{eq:trick}), the
moment of inertia can be decomposed into the valence and sea parts:
\FL
\begin{eqnarray}
\Theta &=& \Theta_{val.} + \Theta_{sea} , \nonumber \\
\Theta_{val.} &=& \mbox{$1\over 2$} N_c \!\!\! \sum_{\begin{array}{c}
                              \scriptstyle i \in  {\rm val.} \\
                              \scriptstyle j \in  {\rm all}
                             \end{array} } \!\!\! {}^{'} \,
 \frac{{\left | \langle i | \tau_3 | j \rangle \right |}^2}
 {\varepsilon_i - \varepsilon_j} , \;\; \Theta_{sea} = \mbox{$1\over 2$}
N_c \!\!\!\!\!\! \sum_{\begin{array}{c}
                              \scriptstyle i \in  {\rm sea} \\
                              \scriptstyle j \in  {\rm pos. en.}
                             \end{array} } \!\!\!\!\!
 \frac{{\left | \langle i | \tau_3 | j \rangle \right |}^2}
 {\varepsilon_i - \varepsilon_j} . \nonumber \\
\label{eq:thetavalsea}
\end{eqnarray}
Note that the ``full'' expression (\ref{eq:thetaspectral}) obeys
the Pauli exclusion principle, hence using
Eq. (\ref{eq:trick}) we have broken the the original expression into
two parts, each of which violates the Pauli principle.
In fact, an analogous decomposition is used in the treatment of the
relativistic fermion propagator in fermion matter \cite{Chin}. Below we
explain why this is useful.
Firstly, the expression for $\Theta_{val}$ corresponds to our quark
part of the moment of inertia calculated in Sec. \ref{se:crank}. Secondly,
the sea part of the moment of inertia can be simply approximated
only if it is written as in Eq. (\ref{eq:thetavalsea}). Indeed, we can
write down
\begin{equation}
\Theta_{sea} = \frac{\dot{\imath}}{4} N_c \int \frac{d \omega}{2 \pi}
Sp \frac{1}{\omega - h} \tau_3 \frac{1}{\omega - h} \tau_3 ,
\label{eq:thetaseacont}
\end{equation}
where the contour of integration is given in Fig. \ref{fi:contsea}. This
contour can be Wick-rotated without picking up any pole contributions, and
we obtain an expression in Euclidean space. We can then perform
standard gradient expansion methods \cite{Aitchison,Nepomechi,Zuk,DPPob88}
to rewrite $\Theta_{sea}$ as
an integral over the classical pion field. The first term, with
no derivatives, is just our expression for the pion part of the moment
of inertia, Eq. (\ref{eq:theta}). Furthermore, this term
does not depend on the NJL cut-off, since the normalization factor
is the same as in the pion wave function normalization \cite{NJL}.
Further terms in the gradient expansion do depend on the cut-off.
If we tried to perform the Wick rotation on the original expression
(\ref{eq:thetaNJL}), we would pick up a pole contribution from the
occupied valence level, and our final expression (\ref{eq:thetavalsea})
would also follow.

\newpage
\widetext

\figure{Contour of integration, {\cal C}, for the total
(sea- and valence-quark) contribution:
         {\cal C} cannot be Wick-rotated without picking
up the valence quark contribution.
         Notation for various labels used in the text is visualized.
   \label{fi:contval}}

\figure{Contour of integration for the sea-quark contribution:
{\cal C} can be Wick-rotated to the contour {\cal C}'.
Upon bosonization, the sea-quark
effects can be described by mesonic degrees of freedom.
   \label{fi:contsea}}

\newpage

\mediumtext
\begin{table}
\caption{Matrix elements of {\boldmath $\tau \cdot \widehat{r}$}}
\begin{tabular}{l|cccc}
 & $|K,0 >$ & $|K,1 >$ & $|K-1,1 >$ & $| K+1,1 >$\\
\hline
$<K,0|$ & 0 & 0 & $\sqrt{\frac{K}{2 K+1}}$ & $- \sqrt{\frac{K+1}{2 K+1}}$ \\
$<K,1 |$ & 0 & 0 & $-\sqrt{\frac{K+1}{2 K+1}}$ & $-\sqrt{\frac{K}{2 K+1}}$ \\
$<K-1,1 |$ & $\sqrt{\frac{K}{2 K+1}}$ & $-\sqrt{\frac{K+1}{2 K+1}}$ & 0 & 0 \\
$<K+1,1 |$ & $-\sqrt{\frac{K+1}{2 K+1}}$ & $-\sqrt{\frac{K}{2 K+1}}$ & 0 & 0 \\
\end{tabular}
\label{ta:taudotr}
\end{table}

\begin{table}
\caption{Matrix elements of {\boldmath $\sigma \cdot \widehat{r}$}}
\begin{tabular}{l|cccc}
& $|K,0 >$ & $|K,1 >$ & $|K-1,1 >$ & $| K+1,1 >$ \\
\hline
$<K,0 |$ & 0 & 0 & $-\sqrt{\frac{K}{2 K+1}}$ & $\sqrt{\frac{K+1}{2 K+1}}$ \\
$<K,1 |$ & 0 & 0 & $-\sqrt{\frac{K+1}{2 K+1}}$ & $-\sqrt{\frac{K}{2 K+1}}$ \\
$<K-1,1 |$ & $-\sqrt{\frac{K}{2 K+1}}$ & $-\sqrt{\frac{K+1}{2 K+1}}$ & 0 & 0 \\
$<K+1,1 |$ & $\sqrt{\frac{K+1}{2 K+1}}$ & $-\sqrt{\frac{K}{2 K+1}}$ & 0 & 0 \\
\end{tabular}
\label{ta:sigmadotr}
\end{table}

\begin{table}
\caption{Matrix elements of {\boldmath $\sigma \cdot L$}}
\begin{tabular}{l|cccc}
 & $|K,0 >$ & $|K,1 >$ & $|K-1,1 >$ & $| K+1,1 >$ \\
\hline
$<K,0 |$ & 0 & $-\sqrt{K(K+1)}$ & 0 & 0 \\
$<K,1 |$ & $-\sqrt{K(K+1)}$ & -1 & 0 & 0 \\
$<K-1,1 |$ & 0 & 0 & $K-1$ & 0 \\
$<K+1,1 |$ & 0 & 0 & 0 & $-K-2$ \\
\end{tabular}
\label{ta:sigmadotL}
\end{table}

\narrowtext

\end{document}